# Two-dimensional superconductor-insulator quantum phase transitions in an electron-doped cuprate

S. W. Zeng[1,2], Z. Huang[1], W. M. Lv[1], N. N. Bao[1,3], K. Gopinadhan[1], L. K. Jian[1], T. S. Herng[3], Z. Q. Liu[4], Y. L. Zhao[1], C. J. Li[1], H. J. Harsan Ma[1,2], P. Yang[5], J. Ding[1,3], T. Venkatesan[1,2,6,*], Ariando[1,2,*]

[1]NUSNNI-NanoCore, National University of Singapore, Singapore 117411

[2]Department of Physics, National University of Singapore, Singapore 117542

[3]Department of Materials Science and Engineering, National University of Singapore, Singapore 117576

[4]Oak Ridge National Laboratory, Oak Ridge, Tennessee, United States, 37831

[5]Singapore Synchrotron Light Source (SSLS), National University of Singapore, 5 Research Link, Singapore 117603

[6]Department of Electrical and Computer Engineering, National University of Singapore, Singapore, Singapore 117576

*To whom correspondence should be addressed.

E-mail: venky@nus.edu.sg, ariando@nus.edu.sg

**We use ionic liquid-assisted electric field effect to tune the carrier density in an electron-doped cuprate ultrathin film and cause a two-dimensional superconductor-insulator transition (SIT). The low upper critical field in this system allows us to perform magnetic field ($B$)-induced SIT in the liquid-gated superconducting film. Finite-size scaling analysis indicates that SITs induced both by electric and magnetic field are quantum phase transitions and the transitions are governed by percolation effects - quantum mechanical in the former and classical in the latter case. Compared to the hole-doped cuprates, the SITs in electron-doped system occur at critical sheet resistances ($R_c$) much lower than the pair quantum resistance $R_Q=h/(2e)^2=6.45$ k$\Omega$, suggesting the possible existence of fermionic excitations at finite temperature at the insulating phase near SITs.**

Electric field effect doping has long been a key technology to tune the charge carrier of a material in a reversible, quasi-continuous way and without inducing structural changes [1-5]. More recently, field effect in electronic double layer transistor (EDLT) configuration which use ionic liquids (ILs) and polymer electrolyte as gate dielectrics has been shown to induce large amount of carriers (up to a level of $10^{15}/cm^2$) at the surface of a thin film [6], and thus is attracting growing interest. The capability of EDLTs in accumulating charge carriers has been demonstrated by gate-induced phase transitions in various materials [6-18]. In particular, employing EDLTs superconductivity has been induced in non-superconducting samples such as $SrTiO_3$ [7], $KTaO_3$ [9], $ZrNCl$ [8], $MoS_2$ [15], and hole-doped cuprate $La_{2-x}Sr_xCuO_4$ (LSCO) [11].

Tuning the carrier density in high-$T_c$ cuprates is of particular interest since the properties of cuprates depend dramatically on the charge carriers. Doping both holes and electrons could cause the change of cuprates from antiferromagnetic Mott insulators to high-$T_c$ superconductors. However, what happens near the critical point where superconductor-insulator transition (SIT) occurs, and what are the differences between hole- and electron-doped cuprates at the critical point are key open questions. To address such questions, quasi-continuous tuning of carrier density by electric field effect across the SIT point is required. In hole-doped cuprates LSCO and $YBa_2Cu_3O_{7-x}$ (YBCO), EDLT-tuned two-dimensional (2D) SIT occurred at pair quantum resistance $R_Q$=6.45 kΩ [11, 12], which suggests that Cooper pairs preserve at both superconducting and insulating sides near SIT as opposed to the formation of fermionic excitations. Further, it has been suggested that the SITs in hole-doped system are 2D quantum phase transitions (QPTs) via finite size scaling analysis. However, a counterpart of LSCO, electron-doped $R_{2-x}Ce_xCuO_4$ (R=Pr, Nd, Sm or Eu) exhibit different crystalline structures, phase diagrams and electronic properties [19]. These suggest that SITs in electron-doped cuprates may be expected to be different. Understanding the nature of carrier-induced SIT in electron-doped systems should shed more light on origins of superconductivity in cuprates.

Further, as the SIT strongly depends also on the dynamics of the vortices, it would be revealing if *B*-induced SIT could be studied, as one suppresses the superconductivity via vortices as opposed to charge injection. *B*-induced SITs in cuprates have been investigated in chemically doped superconductors [20, 21]. Unlike chemical doping, superconductivity induced by EDLT has a distinct 2D nature, due to the short screening length in cuprates [11, 22]. *B*-induced SIT in such EDLT superconducting state would be very interesting. However, this has not been done in hole-doped cuprates due to the huge upper critical field $H_{c2}$. In electron-doped cuprates $H_{c2}$ is much lower and this provides an opportunity to study *B*-induced SIT in superconducting EDLT.

In this paper, we simultaneously study the SITs induced by both electric and magnetic fields in electron-doped $Pr_{2-x}Ce_xCuO_4$ (PCCO) ultrathin films. Using EDLT configuration, electrostatically induced superconducting transition in an initially insulating PCCO ultrathin film is observed. In such superconducting EDLTs with different doping levels, insulating states are realized by applying magnetic fields. We found that the SITs induced by both electric and magnetic fields occurred at $R_c$ much lower than $R_Q$. Moreover, finite size scaling analysis is performed and it is suggested that the transitions are 2D-QPTs.

Ultrathin layering structures with nominal 1 unit cell (uc) underdoped PCCO on 4-uc $Pr_2CuO_4$ (PCO) were grown on $TiO_2$-terminated $SrTiO_3$ (001) substrates by pulsed laser deposition (PLD) system. The Ce contents of the top $Pr_{2-x}Ce_xCuO_4$ layers are $x_{Ce}$=0.1 (sample A, B) and 0.04 (sample C). The thin films were deposited at 790 $^o$C under oxygen partial pressure ($P_{O2}$) of 0.25 mbar and then cooled down to room temperature from 720 $^o$C in vacuum ($P_{O2}$<10$^{-4}$ mbar) at a cooling rate of 20 $^o$C/min. Fig. 1(a) shows atomic force microscopy (AFM) image of ultrathin PCCO/PCO films. The root-mean-square (RMS) roughness is ~0.32 nm, indicating smooth surface of thin films. Devices with a Hall bar geometry were fabricated to accurately measure the sheet resistance. Before deposition of thin films, $SrTiO_3$ substrates were patterned into Hall bar geometry by using conventional photolithography and depositing amorphous AlN films as hard masks. Then, the patterned substrates were put into PLD chamber for the deposition of thin films. Finally, after film deposition, Cr/Au (10/70 nm) layers were deposited for current/voltage probes and gate electrodes. A micrograph of a fabricated device is shown in Fig. 1(b). The width of the Hall bar is 50 μm and length is 500 μm. Six-probe configuration of the Hall bar allows for the measurement of both longitudinal and Hall resistance. A large planar gate electrode is used to accumulate ions.

A small droplet of the IL, N,N-diethyl-N-methyl-N-(2-methoxyethyl) ammonium bis(trifluoromethyl sulphonyl)imide (DEME-TFSI), covered both the conducting channel and the gate electrode. Then the sample was put into the chamber of a Quantum Design Physical Property Measurement System (PPMS) for the transport measurements. The gate voltage ($V_G$) was applied at 210 K and kept for 20~40 min for charging. Resistance measurement was made as the samples were cooled down while keeping the $V_G$ constant. In order to change $V_G$, the sample was heated to 210 K after each resistance-temperature curve measurement and a new $V_G$ was applied. As is shown in Fig. 1(c), when a positive $V_G$ is applied, the mobile cations accumulate on the transport channel and induce electrons on the surface of the thin film. Fig. 1(d) shows the leakage current ($I_G$) as a function of $V_G$ for a PCCO-EDLT device. It can be seen that $I_G$ is on the order of 1 nA. This negligibly small $I_G$ indicates good operation of device. In these experiments the IL predominantly induces charges at the surface as opposed to changing the local oxygen vacancy concentration [17, 23].

Figures 2(a)-(b) show the sheet resistance versus temperature ($R_s$-$T$) curves at various $V_G$ for samples with $x_{Ce}$=0.1 (sample A) and $x_{Ce}$=0.04 (sample C). The $R_s$-$T$ curves of sample B can be seen in Supplemental Material Fig. S8 [24]. Since the Ce content is lower in sample C, initial resistance at $V_G$=0 V is much higher than that of sample A.

Many curves were recorded and insulator-to-superconductor transitions were observed in both of the samples. The initially underdoped samples show insulating behavior. Even at the $V_G$ near SIT, the resistance increases with decreasing $T$ below ~140 K, suggesting that EDLT samples have a much higher resistance at the insulating phase (Supplemental Material Section 6) [24]. As $V_G$ is increased, the accumulation of electrons on the surface is enhanced and the $R_s$ decreases. Note that the bottom PCO layer does not contribute to this resistance drop as verified by direct field effect experiments on PCO without a top PCCO layer. At $V_G$ above certain values (2.59 V for sample A and 2.41 V for sample C), the resistance drops sharply at low $T$, signaling the onset of superconductivity. As $V_G$ is increased, the $T_c$ increases up to a maximum of ~14 K and then decreases for further increasing $V_G$. The $T_c$ is taken to be the $T$ at which the $R_s$ falls to 50% of its normal-state value. Due to the short Thomas-Fermi screening length in cuprates [11, 22], the active layer in the EDLT is limited to one or two $CuO_2$ planes, leaving the deeper layers unaffected. To further demonstrate this screening effect, devices with optimally doped $Pr_{1.85}Ce_{0.15}CuO_4$ showing superconducting $T_c$ of ~15 K were fabricated and tested. It was found that even at a higher $V_G$ of 4 V, only a small change of $T_c$ but no insulating phase was obtained in a film with thickness above 3 uc, indicating the shunting effect from the deeper layers which are not influenced by the field effect. These suggest that the superconductivity obtained in the underdoped EDLT devices occurs within one or two $CuO_2$ layers on the surface, and thus, is two-dimensional. This is different from the multilayers of electron-doped cuprates in which the redistribution of charge could be over a large distance, and thus, the transport properties of one layer could be affected by other ones [25].

The electric field effect-induced phase diagram of $T_c$ versus carrier concentration, $x$, is constructed based on the measured $R$-$T$ curves at different $V_G$ and shown in Fig. 3. The details of the estimation of $x$ can be seen in Supplemental Material Section 3[26]. For comparison, field effect-induced phase diagram of hole-doped LSCO is extracted from the reference [11] and plotted. The superconductivity occurs at doping levels of $x \approx 0.12$ for sample A, $x \approx 0.11$ for sample B and C, comparable to that in chemically doped PCCO [27]. These doping levels are much higher than that of LSCO, indicating that more charge carriers are required to induce superconductivity in electron-doped cuprate EDLT. This could also be seen from the fact that the $V_G$ at which superconductivity occurs in PCCO (above 2 V) is higher than that (1.25 V) in LSCO [11]. Inset of Fig. 3 shows the normalized $T_c$, $T_c/T_c(x=0.15)$, as a function of $x$ for field effect and chemical doping [27]. One can see that the phase diagrams derived from field effect and chemical doping are reasonably similar. This suggests that the estimation of $x$ is reliable, a necessity for quantitative analysis near the SIT. The highest $T_c$ obtained by field effect doping is ~14 K which is lower than that of LSCO (~29 K), and that (~21 K) obtained by chemical doping [27]. The difference of $T_c$ between EDLTs and chemically doped samples has also been observed in hole-doped LSCO [11]. This may be due to the possible disorder in the ultrathin film as well as the 2D nature of superconducting EDLTs.

It is proposed that SIT at the limit of zero $T$ and 2D is an example of a QPT [28-31]. At nonzero $T$, the signature of a QPT is a success of finite size scaling in describing the finite-$T$ data [2, 11, 12, 28-34]. For a 2D system, the $R_s$ near a quantum critical point collapses onto a single scaling function, $R_s(x_t,T)=R_cF(|x_t-x_c|T^{-1/\nu z})$, where $T$ is temperature, $x_t$ is tuning parameter (in the present result, $x_t$ is carrier concentration and magnetic field), $x_c$ is $x_t$ at critical point, $R_c$ is critical $R_s$ at $x_t=x_c$, $\nu$ is the correlation length critical exponent, $z$ is the dynamic critical exponent, and $F(u)$ is a universal function of $u$ with $F(u)\to 1$ when $u\to 0$. The low-$T$ resistances near SIT were extracted for quantitative analysis. Fig. 4 shows the resistance isotherms of the sample C from 2.2 K to 7 K as a function of $x$. One can see that all curves cross at a single point which separates the insulating and superconducting regimes. The data at the point give the critical values of $x$ and $R_s$ to be $x_c \approx 0.1005$ and $R_c \approx 2.88$ k$\Omega$. The $R_c$ is much lower than $R_Q$. Inset of Fig. 4 is the finite size scaling analysis of the data in the vicinity of SIT. Given the exponent product $\nu z = 2.4$, all of the isotherm curves collapse onto a single function, suggesting the occurrence of a 2D-QPT. The scaling analysis of other samples is shown in Supplemental Material Fig. S7-S9 [24], and the similar values of $R_c$ and $\nu z$ are also observed, indicating that our results are reproducible. The value of $\nu z$ for PCCO is different from those for the hole-doped 214-structure counterpart, $\nu z$=1.5 for LSCO [11] and $\nu z$=1.2 for La$_2$CuO$_{4+\delta}$ [13]. It is consistent with that of the quantum percolation model ($\nu z \approx 7/3$) [12, 35-38].

Magnetic field ($B$)-tuned SITs were done on sample A at three different $V_G$ of 2.75, 3.0 and 3.2 V, which correspond to underdoped, optimally doped and overdoped states, respectively. At each $V_G$ the sample was initially superconducting and could be tuned to insulating states (Supplemental Material Fig. S6) [24]. Figure 5(a)-(c) show the $R_s$ as a function of $B$ at four $T$ points below the onset of superconductivity. At each doping level, isotherms of $R_s$ cross each other at a single point which separates the insulating from superconducting phase. The critical magnetic fields, $B_c$ and $R_c$ as a function of $x$ are shown in Fig. 5(d). It can be seen that $B_c$ peaks at optimally doped state and shows lower values at underdoped and overdoped states. $R_c$ is similar to $R_n$ at each doping level, decreases with increasing doping levels and is much lower than $R_Q$, especially for the ones at optimally doped and overdoped states. The insets of Fig. 5(a)-5(c) show finite size scaling analysis of isotherm curves. One can see that all the curves collapse onto a single function except at high $B$. The deviation of scaling at high $B$ could probably be due to the presence of weak localization dominating at high $B$, which causes a weak increase in resistance with increasing $B$ [39]. This can be seen in the $R_s$-$B$ curves, which show that above $B_c$, the resistance tends to saturate at high $B$. Interestingly, the same exponent product $vz$=1.4 is obtained at three different $V_G$. The $vz$ observed here is consistent with classical percolation theory ($vz$≈4/3) [32, 37, 40] and is similar to that of the chemically doped sample [21]. Note that for the case of chemical doping, the $B$-tuned SIT has only been observed in the underdoped sample. In our EDLTs, $B$-tuned SITs occurred at different doping states, and the same $vz$ was observed. This suggests that $B$-tuned SITs in EDLT are governed by the same mechanism at different doping levels.

In the bosonic picture of SIT the perfect duality between Cooper pairs and vortices predicts that $R_c$ is equal to $R_Q = h/(2e)^2 = 6.45$ kΩ [11, 28, 30, 31, 35, 41, 42]. The values of $R_c \approx R_Q$ observed in YBCO and LSCO suggest that the SITs are driven by quantum phase fluctuations in the hole-doped cuprates, consistent with the bosonic picture [11, 12]. However, in electron-doped PCCO we observed $R_c$ much lower than $R_Q$ in both carrier- and $B$-tuned SITs. This suggests that there are possible fermionic excitations at finite $T$ contributing to the conduction near SIT in PCCO [32]. Further, it is found that in the $B$-induced SIT, $R_c$ shares the same evolution with $R_n$ and decreases with increasing charge carriers (Fig. 5(d)). Moreover, as is shown in electric field effect-induced phase diagram (Fig. 3), the carrier density at which SIT occurs in PCCO is higher than that in hole-doped ones. These observations suggest that fermionic excitations at the insulating side of SIT may occur in n-type superconducting cuprate phases possibly on account of the higher carrier density compared to the hole-doped system. However, to fully understand the role of fermionic excitations on the SIT of electron-doped cuprates further studies will be needed [31, 43, 44].

In summary, we investigated carrier- and $B$-tuned SITs in electron-doped PCCO ultrathin films. Using IL as the dielectric material in EDLTs, the initially insulating states of underdoped PCCO could be gated into superconducting states. In superconducting EDLTs, insulating states were realized by applying $B$. Through finite-size scaling analysis, it was suggested that carrier- and $B$-tuned SITs are 2D-QPTs. Moreover, we found the $R_c$ in PCCO are much lower than $R_Q$, which probably suggests that there are fermionic excitations at finite $T$ contributing to the conduction near SIT. The present results could help to further our understanding of quantum criticalities in cuprates.

**Acknowledgments:**

This work is supported by the Singapore National Research Foundation (NRF) under the Competitive Research Programs (CRP Award No. NRF-CRP 8-2011-06 and CRP Award No. NRF-CRP10-2012-02) and the NUS FRC (AcRF Tier 1 Grant No. R-144-000-346-112). P. Yang is supported by the SSLS via NUS Core Support C-380-003-003-001.

[1] C. H. Ahn, A. Bhattacharya, M. Di Ventra, J. N. Eckstein, C. D. Frisbie, M. E. Gershenson, A. M. Goldman, I. H. Inoue, J. Mannhart, A. J. Millis, A. F. Morpurgo, D. Natelson, J. M. Triscone, *Electrostatic modification of novel materials*, Rev. Mod. Phys., **78**, 1185 (2006).

[2] K. A. Parendo, K. H. Sarwa, B. Tan, A. Bhattacharya, M. Eblen-Zayas, N. E. Staley, A. M. Goldman, *Electrostatic tuning of the superconductor-insulator transition in two dimensions*, Phys. Rev. Lett., **94**, 197004 (2005).

[3] D. Matthey, N. Reyren, J. M. Triscone, T. Schneider, *Electric-field-effect modulation of the transition temperature, mobile carrier density, and in-plane penetration depth of NdBa2Cu3O7-delta thin films*, Phys. Rev. Lett., **98**, 057002 (2007).

[4] A. D. Caviglia, S. Gariglio, N. Reyren, D. Jaccard, T. Schneider, M. Gabay, S. Thiel, G. Hammerl, J. Mannhart, J. M. Triscone, *Electric field control of the LaAlO3/SrTiO3 interface ground state*, Nature, **456**, 624 (2008).

[5] X. X. Xi, C. Doughty, A. Walkenhorst, C. Kwon, Q. Li, T. Venkatesan, *Effects of Field-Induced Hole-Density Modulation on Normal-State and Superconducting Transport in YBa2Cu3O7-x*, Phys. Rev. Lett., **68**, 1240 (1992).

[6] H. T. Yuan, H. Shimotani, A. Tsukazaki, A. Ohtomo, M. Kawasaki, Y. Iwasa, *High-Density Carrier Accumulation in ZnO Field-Effect Transistors Gated by Electric Double Layers of Ionic Liquids*, Advanced Functional Materials, **19**, 1046 (2009).

[7] K. Ueno, S. Nakamura, H. Shimotani, A. Ohtomo, N. Kimura, T. Nojima, H. Aoki, Y. Iwasa, M. Kawasaki, *Electric-field-induced superconductivity in an insulator*, Nature Materials, **7**, 855 (2008).

[8] J. T. Ye, S. Inoue, K. Kobayashi, Y. Kasahara, H. T. Yuan, H. Shimotani, Y. Iwasa, *Liquid-gated interface superconductivity on an atomically flat film*, Nature Materials, **9**, 125 (2010).

[9] K. Ueno, S. Nakamura, H. Shimotani, H. T. Yuan, N. Kimura, T. Nojima, H. Aoki, Y. Iwasa, M. Kawasaki, *Discovery of superconductivity in KTaO3 by electrostatic carrier doping*, Nature Nanotechnology, **6**, 408 (2011).

[10] Y. Yamada, K. Ueno, T. Fukumura, H. T. Yuan, H. Shimotani, Y. Iwasa, L. Gu, S. Tsukimoto, Y. Ikuhara, M. Kawasaki, *Electrically Induced Ferromagnetism at Room Temperature in Cobalt-Doped Titanium Dioxide*, Science, **332**, 1065 (2011).

[11] A. T. Bollinger, G. Dubuis, J. Yoon, D. Pavuna, J. Misewich, I. Bozovic, *Superconductor-insulator transition in La2-xSrxCuO4 at the pair quantum resistance*, Nature, **472**, 458 (2011).

[12] X. Leng, J. Garcia-Barriocanal, S. Bose, Y. Lee, A. M. Goldman, *Electrostatic Control of the Evolution from a Superconducting Phase to an Insulating Phase in Ultrathin YBa2CaCu3O7-x Films*, Phys. Rev. Lett., **107**, 027001 (2011).

[13] J. Garcia-Barriocanal, A. Kobrinskii, X. Leng, J. Kinney, B. Yang, S. Snyder, A. M. Goldman, *Electronically driven superconductor-insulator transition in electrostatically doped La2CuO4+delta thin films*, Phys. Rev. B, **87**, 024509 (2013).

[14] A. S. Dhoot, S. C. Wimbush, T. Benseman, J. L. MacManus-Driscoll, J. R. Cooper, R. H. Friend, *Increased Tc in Electrolyte-Gated Cuprates*, Advanced Materials, **22**, 2529 (2010).

[15] J. T. Ye, Y. J. Zhang, R. Akashi, M. S. Bahramy, R. Arita, Y. Iwasa, *Superconducting Dome in a Gate-Tuned Band Insulator*, Science, **338**, 1193 (2012).

[16] M. Nakano, K. Shibuya, D. Okuyama, T. Hatano, S. Ono, M. Kawasaki, Y. Iwasa, Y. Tokura, *Collective bulk carrier delocalization driven by electrostatic surface charge accumulation*, Nature, **487**, 459 (2012).

[17] J. Jeong, N. Aetukuri, T. Graf, T. D. Schladt, M. G. Samant, S. S. P. Parkin, *Suppression of Metal-Insulator Transition in VO2 by Electric Field-Induced Oxygen Vacancy Formation*, Science, **339**, 1402 (2013).

[18] X. Leng, J. Garcia-Barriocanal, J. Kinney, B. Y. Yang, Y. Lee, A. M. Goldman, *Electrostatic tuning of the superconductor to insulator transition of YBa2Cu3O7-x using ionic liquids*, Journal of Physics: Conference Series **449**, 012009 (2013).

[19] N. P. Armitage, P. Fournier, R. L. Greene, *Progress and perspectives on electron-doped cuprates*, Rev. Mod. Phys., **82**, 2421 (2010).

[20] G. T. Seidler, T. F. Rosenbaum, B. W. Veal, *2-dimensional superconductor-insulator transition in bulk single-crystal YBa2Cu3O6.38*, Phys. Rev. B, **45**, 10162 (1992).

[21] S. Tanda, S. Ohzeki, T. Nakayama, *Bose-glass vortex-glass phase-transition and dynamic scaling for high-Tc Nd2-xCexCuO4 thin-films*, Phys. Rev. Lett., **69**, 530 (1992).

[22] S. Smadici, J. C. T. Lee, S. Wang, P. Abbamonte, G. Logvenov, A. Gozar, C. D. Cavellin, I. Bozovic, *Superconducting Transition at 38 K in Insulating-Overdoped La2CuO4-La1.64Sr0.36CuO4 Superlattices: Evidence for Interface Electronic Redistribution from Resonant Soft X-Ray Scattering*, Phys. Rev. Lett., **102**, 107004 (2009).

[23] See Supplemental Material Section 2 for testing device reversibility.

[24] See Supplemental Material Section 4 for measurement of *R-T* curves at various magnetic field, Section 5 for testing reproducibility of EDLT experiments, and Section 6 for the discussion of insulating behavior near SIT.

[25] K. Jin, P. Bach, X. H. Zhang, U. Grupel, E. Zohar, I. Diamant, Y. Dagan, S. Smadici, P. Abbamonte, R. L. *Greene, Anomalous enhancement of the superconducting transition temperature of electron-doped La2-xCexCuO4 and Pr2-xCexCuO4 cuprate heterostructures*, Phys. Rev. B, **83**, 060511(2011).

[26] See Supplemental Material Section 3 for detailed estimation of carrier concentration.

[27] J. L. Peng, E. Maiser, T. Venkatesan, R. L. Greene, G. Czjzek, *Concentration range for superconductivity in high-quality Pr2-xCexCuO4-y thin films*, Phys. Rev. B, **55**, R6145 (1997).

[28] M. P. A. Fisher, *Quantum phase-transitions in disordered 2-dimensional superconductors*, Phys. Rev. Lett., **65**, 923 (1990).

[29] S. L. Sondhi, S. M. Girvin, J. P. Carini, D. Shahar, *Continuous quantum phase transitions*, Rev. Mod. Phys., **69**, 315 (1997).

[30] V. F. Gantmakher, V. T. Dolgopolov, *Superconductor-insulator quantum phase transition*, Physics-Uspekhi, **53**, 1 (2010).

[31] A. M. Goldman, N. Markovic, *Superconductor-insulator transitions in the two-dimensional limit*, Physics Today, **51**, 39 (1998).

[32] A. Yazdani, A. Kapitulnik, *Superconducting-insulating transition in 2-dimensional alpha-moge thin-films*, Phys. Rev. Lett., **74**, 3037 (1995).

[33] Y. Liu, K. A. McGreer, B. Nease, D. B. Haviland, G. Martinez, J. W. Halley, A. M. Goldman, *Scaling of the insulator-to-superconductor transition in ultrathin amorphous Bi films*, Phys. Rev. Lett., **67**, 2068 (1991).

[34] N. Markovic, C. Christiansen, A. M. Mack, W. H. Huber, A. M. Goldman, *Superconductor-insulator transition in two dimensions*, Phys. Rev. B, **60**, 4320 (1999).

[35] M. A. Steiner, N. P. Breznay, A. Kapitulnik, *Approach to a superconductor-to-Bose-insulator transition in disordered films*, Phys. Rev. B, **77**, 212501 (2008).

[36] D. H. Lee, Z. Q. Wang, S. Kivelson, *Quantum percolation and plateau transitions in the quantum hall-effect*, Phys. Rev. Lett., **70**, 4130 (1993).

[37] Y. Dubi, Y. Meir, Y. Avishai, *Unifying model for several classes of two-dimensional phase transition*, Phys. Rev. Lett., **94**, 156406 (2005).

[38] R. Schneider, A. G. Zaitsev, D. Fuchs, H. von Lohneysen, *Superconductor-Insulator Quantum Phase Transition in Disordered FeSe Thin Films*, Phys. Rev. Lett., **108**, 257003 (2012).

[39] A. Kussmaul, J. S. Moodera, P. M. Tedrow, A. Gupta, *2-dimensional character of the magnetoresistance in $Nd_{1.85}Ce_{0.15}CuO_{4-\delta}$ thin-films*, Physica C, **177**, 415 (1991).

[40] N. Mason, A. Kapitulnik, *Dissipation effects on the superconductor-insulator transition in 2D superconductors*, Phys. Rev. Lett., **82**, 5341 (1999).

[41] D. B. Haviland, Y. Liu, A. M. Goldman, *Onset of superconductivity in the two-dimensional limit*, Phys. Rev. Lett., **62**, 2180 (1989).

[42] A. F. Hebard, M. A. Paalanen, *Magnetic-field-tuned superconductor-insulator transition in 2-dimensional films*, Phys. Rev. Lett., **65**, 927 (1990).

[43] S. Dukan, Z. Tesanovic, *Superconductivity in a high magnetic-field - excitation spectrum and tunneling properties*, Phys. Rev. B, **49**, 13017 (1994).

[44] H. Aubin, C. A. Marrache-Kikuchi, A. Pourret, K. Behnia, L. Berge, L. Dumoulin, J. Lesueur, *Magnetic-field-induced quantum superconductor-insulator transition in $Nb_{0.15}Si_{0.85}$*, Phys. Rev. B, **73**, 094521 (2006).

[45] M. Brinkmann, T. Rex, H. Bach, K. Westerholt, *Extended Superconducting Concentration Range Observed in $Pr_{2-x}Ce_xCuO_{4-\delta}$*, Phys Rev Lett, **74**, 4927 (1995).

[46] Y. Krockenberger, J. Kurian, A. Winkler, A. Tsukada, M. Naito, L. Alff, *Superconductivity phase diagrams for the electron-doped cuprates $R(2-x)Ce(x)CuO(4)$ (R=La, Pr, Nd, Sm, and Eu)*, Phys Rev B, **77**, 060505 (2008).

[47] O. Matsumoto, A. Utsuki, A. Tsukada, H. Yamamoto, T. Manabe, M. Naito, *Synthesis and properties of superconducting T'-$R_2CuO_4$ (R=Pr, Nd, Sm, Eu, Gd)*, Phys Rev B, **79**, 100508 (2009).

[48] P. Fournier, J. Higgins, H. Balci, E. Maiser, C.J. Lobb, R.L. Greene, *Anomalous saturation of the phase coherence length in underdoped $Pr_{2-x}Ce_xCuO_4$ thin films*, Phys Rev B, **62**, 11993 (2000).

[49] T. Sekitani, M. Naito, N. Miura, *Kondo effect in underdoped n-type superconductors*, Phys Rev B, **67**, 174503 (2003).

[50] F. Rullier-Albenque, H. Alloul, R. Tourbot, *Disorder and transport in cuprates: Weak localization and magnetic contributions*, Phys Rev Lett, **87**, 157001 (2001).

[51] Y. Dagan, M.C. Barr, W.M. Fisher, R. Beck, T. Dhakal, A. Biswas, R.L. Greene, *Origin of the anomalous low temperature upturn in the resistivity of the electron-doped cuprate superconductors*, Phys Rev Lett, **94**, 057005 (2005).

[52] M. Z. Cieplak, A. Malinowski, S. Guha, M. Berkowski, *Localization and interaction*

*effects in strongly underdoped La2-xSrxCuO4*, Phys Rev Lett, **92**, 187003 (2004).

[53] S. Finkelman, M. Sachs, G. Droulers, N.P. Butch, J. Paglione, P. Bach, R.L. Greene, Y. Dagan, *Resistivity at low temperatures in electron-doped cuprate superconductors*, Phys Rev B, **82**, 094508 (2010).

[54] Y. Onose, Y. Taguchi, K. Ishizaka, Y. Tokura, *Charge dynamics in underdoped Nd(2-x)Ce(x)CuO(4): Pseudogap and related phenomena*, Phys Rev B, **69**, 024504 (2004).

[55] E.M. Motoyama, G. Yu, I.M. Vishik, O.P. Vajk, P.K. Mang, M. Greven, *Spin correlations in the electron-doped high-transition-temperature superconductor Nd2-xCexCuO4 +/-delta*, Nature, 445, 186 (2007).

[56] Y. Dagan, M. M. Qazilbash, C. P. Hill, V. N. Kulkarni, R. L. Greene, *Evidence for a quantum phase transition in Pr2-xCexCuO4-delta from transport measurements*, Phys Rev Lett, **92**, 167001 (2004).

[57] F. F. Balakirev, J. B. Betts, A. Migliori, I. Tsukada, Y. Ando, G. S. Boebinger, *Quantum Phase Transition in the Magnetic-Field-Induced Normal State of Optimum-Doped High-T-c Cuprate Superconductors at Low Temperatures*, Phys Rev Lett, 102, 017004 (2009).

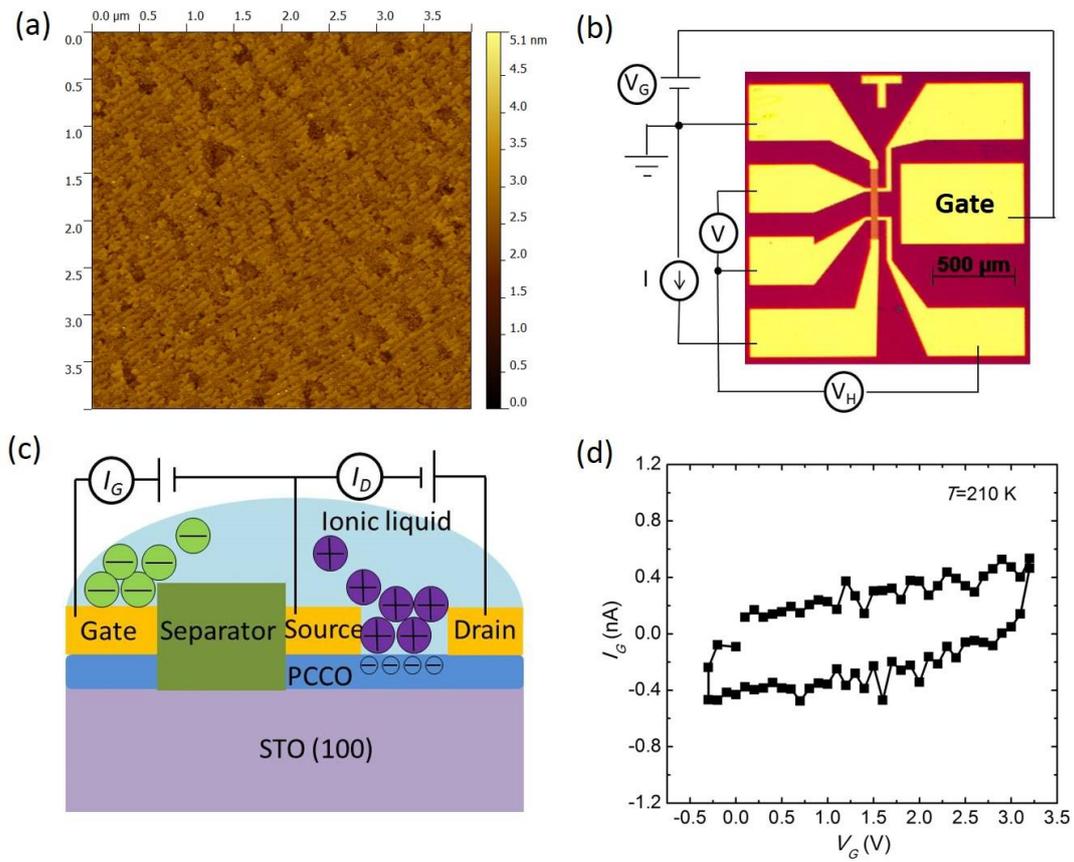

FIG. 1. (a) Atomic force microscopy image (4um×4um) of the surface of ultrathin PCCO/PCO films. (b) Optical micrograph of a typical device and the measurement circuit. (c) Schematic diagram of the operation of an EDLT. (d) Gate voltage ($V_G$) dependence of the leakage current ($I_G$).

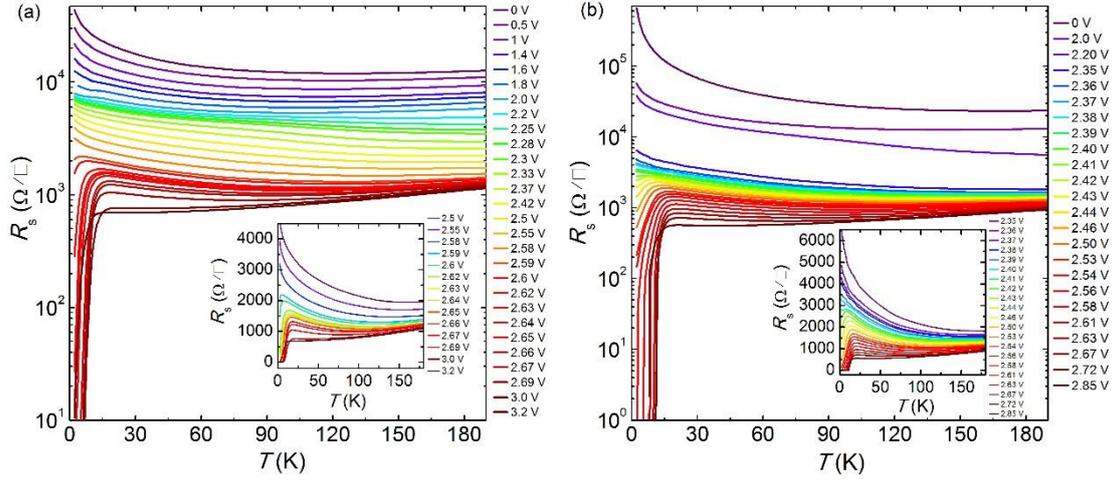

FIG. 2. The logarithmic-scale sheet resistance versus temperature ($R_s$-$T$) curves at various $V_G$ for samples with Ce contents of (a) $x_{Ce}$=0.1 (sample A) and (b) $x_{Ce}$=0.04 (sample C). The insets are the linear-scale $R_s$-$T$ curves near the SITs.

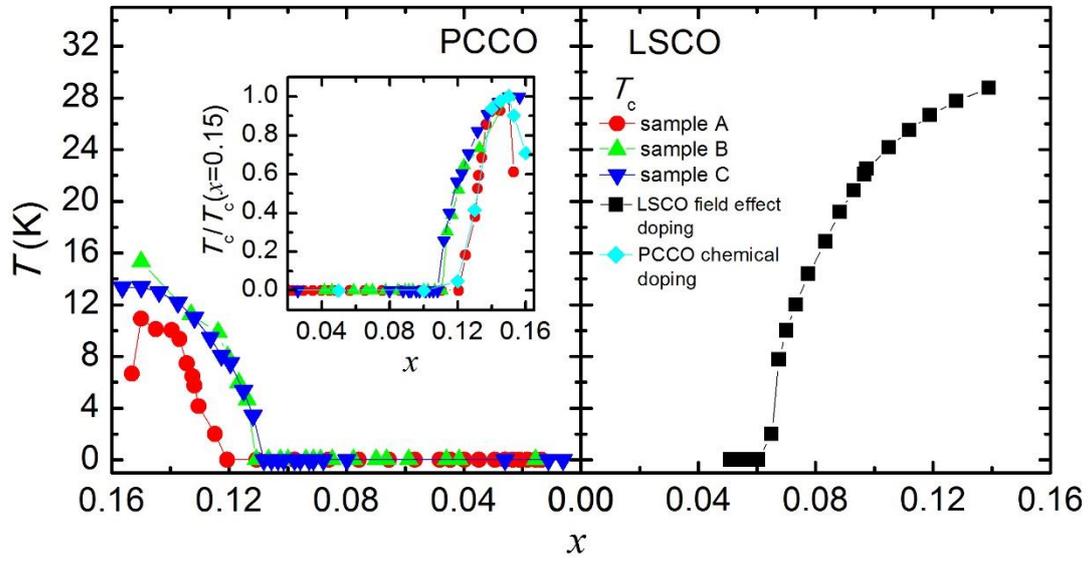

FIG 3. Electric field effect-tuned critical temperature $T_c$ versus carrier concentration $x$ for electron-doped PCCO (sample A, B and C) and hole-doped LSCO. $T_c$ versus $x$ for LSCO is extracted from ref.[11]. Inset shows the normalized $T_c$, $T_c/T_c(x=0.15)$, as a function of $x$ for electric field effect and chemical doping. $T_c/T_c(x=0.15)$ versus $x$ for chemical doping of PCCO is extracted from ref.[27].

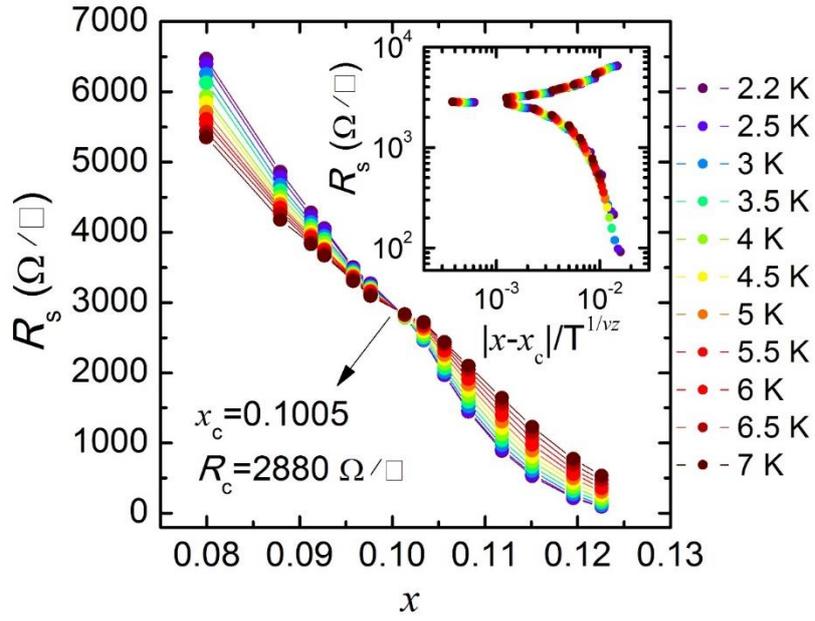

FIG. 4. Isotherms of $R_s$ as a function of $x$ at $T$ ranging from 2.2 to 7 K for sample C. The inset is the finite-size scaling of the isotherm curves, showing the resistance as a function of $|x-x_c|/T^{1/vz}$ with a critical exponent $vz$=2.4.

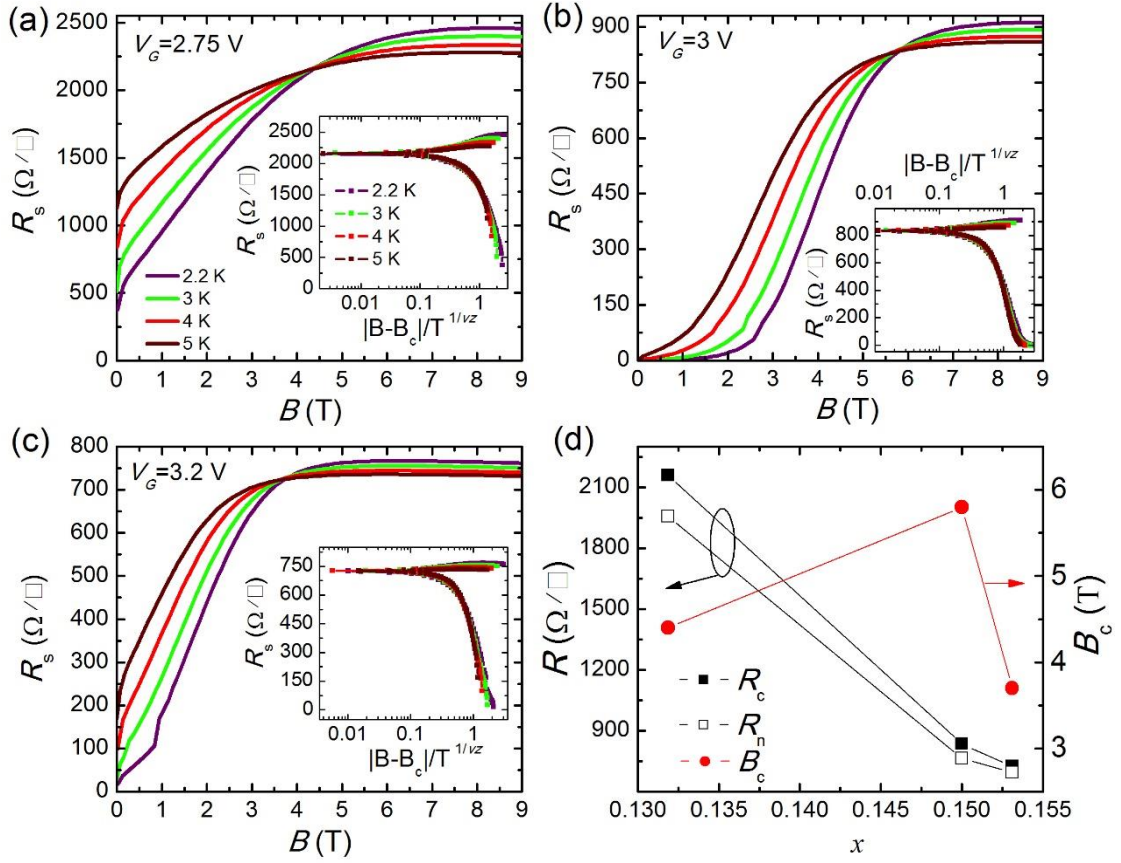

FIG. 5. Resistances of sample A as a function of magnetic fields, $B$, at (a) $V_G$=2.75 V, (b) $V_G$=3.0 V, (c) $V_G$=3.2 V, corresponding to underdoped, optimally doped and overdoped states ($x$=0.132, 0.15 and 0.153), respectively. Insets show the $R_s$ as a function of $|B-B_c|/T^{1/\nu z}$. The data can be fitted by finite size scaling function assuming $\nu z$=1.4 for three doping levels. (d) $R_c$, $R_n$, and $B_c$, as a function of $x$ which correspond to different $V_G$. $R_c$ and $B_c$ are the resistances and magnetic fields at the points where the resistance isotherms cross each other. $R_n$ is the resistance obtained at onset of superconductivity and zero magnetic field (Supplemental Material Fig. S6 (a)).

**Two-dimensional superconductor-insulator quantum phase transitions in an electron-doped cuprate**

*Supplemental Material*

S. W. Zeng[1,2], Z. Huang[1], W. M. Lv[1], N. N. Bao[1,3], K. Gopinadhan[1], L. K. Jian[1], T. S. Herng[3], Z. Q. Liu[4], Y. L. Zhao[1], C. J. Li[1], H. J. Harsan Ma[1,2], P. Yang[5], J. Ding[1,3], T. Venkatesan[1,2,6,*], Ariando[1,2,*]

[1]NUSNNI-NanoCore, National University of Singapore, Singapore 117411

[2]Department of Physics, National University of Singapore, Singapore 117542

[3]Department of Materials Science and Engineering, National University of Singapore, Singapore 117576

[4]Oak Ridge National Laboratory, Oak Ridge, Tennessee, United States, 37831

[5]Singapore Synchrotron Light Source (SSLS), National University of Singapore, 5 Research Link, Singapore 117603

[6]Department of Electrical and Computer Engineering, National University of Singapore, Singapore 117576

*To whom correspondence should be addressed.

E-mail: venky@nus.edu.sg, ariando@nus.edu.sg

## 1. Measurement of X-ray diffraction and reflectivity

In order to further confirm the high quality and smooth surface of thin films, measurements of high-resolution X-ray diffraction (HR-XRD) and low-angle X-ray reflectivity (XRR) were performed. Figure S1 shows the HR-XRD and XRR of a ~45-nm $Pr_{1.9}Ce_{0.1}CuO_4$ film grown on SrTiO3 (001) substrate. These measurements were done in the X-ray Demonstration and Development (XDD) beamline at Singapore Synchrotron Light Source (SSLS). The wavelength of X-ray is λ=1.538 Å. The finite-thickness oscillations and low-angle reflectance oscillations indicate high quality of the film.

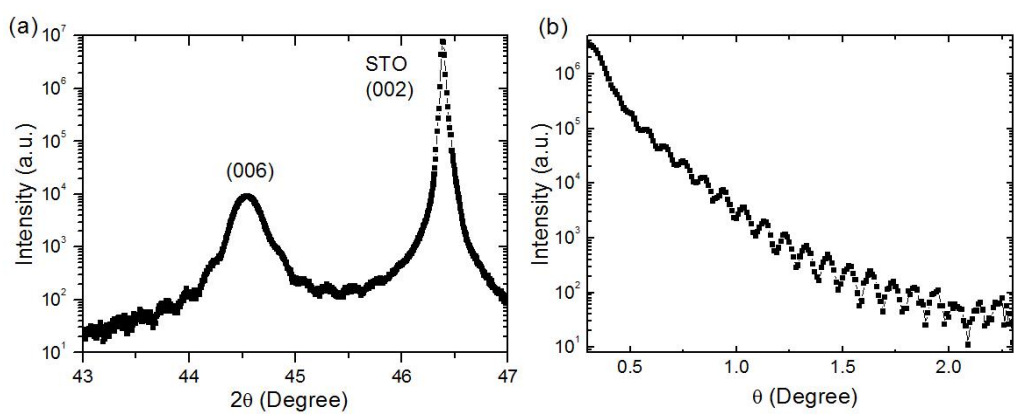

Fig. S1. X-ray data for the $Pr_{1.9}Ce_{0.1}CuO_4$ film. The wavelength of X-ray is 1.538 Å. (a) Finite-thickness oscillations in the vicinity of (006) diffraction peak. (b) Low-angle X-ray reflectance oscillations.

## 2. Reversibility of EDLT operation

The EDLT operation of PCCO was very stable, when the $V_G$ is limited to 3.2 V at 210 K. Figure S2 shows the sheet resistance-temperature ($R_s$-$T$) curves of a device before and after gating experiments. The $R_s$-$T$ curve before gating was obtained on the device in pristine condition before any ionic liquid (IL) was applied. The one after gating was obtained after the device was gated from insulator to superconductor by applying $V_G$ from 0 V up to 3.2 V (inset of Fig. S2), with subsequent removal of the IL. One can see that these two curves show reasonable agreement, indicating the reversibility of EDLT operation. It should be noted that liquid gating could induce oxygen vacancies in some of the oxides, and thereby change the electrical transport properties [17]. However, in our case, the transport properties are remarkably similar before and after gating, suggesting that the chemical compositions are unchanged. Furthermore, the top layer in our EDLT device is underdoped $Pr_{2-x}Ce_xCuO_4$ (x=0.04, 0.1), which is not superconducting even if the oxygen is reduced [27]. However, superconductivity in $Pr_{2-x}Ce_xCuO_4$ is obtained by liquid gating. These indicate that in this study, the IL predominantly induces charges at the surface of PCCO as opposed to inducing oxygen vacancies.

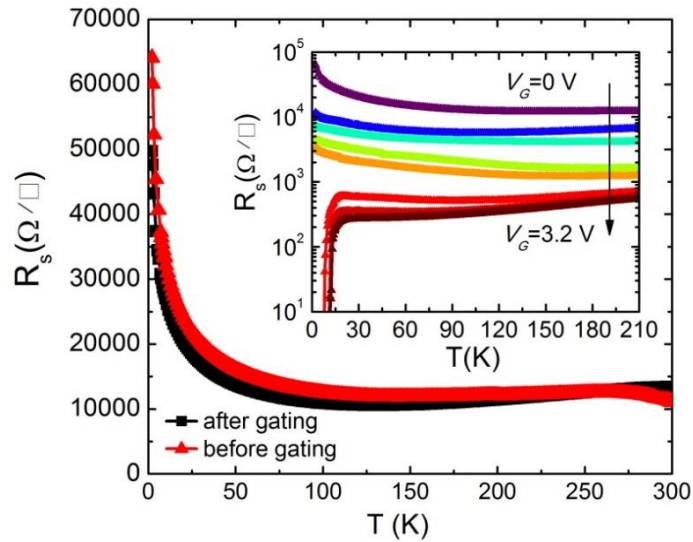

Fig. S2. Sheet resistance-temperature ($R_s$-$T$) curves for the EDLT device before and after gating experiments. Inset: $R_s$-$T$ curves at various $V_G$.

In order to further suggest that the field effect does not change the chemical composition in the thin films, we performed the atomic force microscopy (AFM) measurements on one sample surface before and after gating experiments to investigate the ionic-liquid effect on the sample surface, as is shown in Fig. S3. The root-mean-square (RMS) roughness of the surface before gating experiment is 0.57 nm, which is comparable to that (0.65 nm) of the surface after gating experiment. This further suggests that the ionic-liquid effect on the sample surface is not significant, and thus, the change of chemical composition in PCCO is negligible.

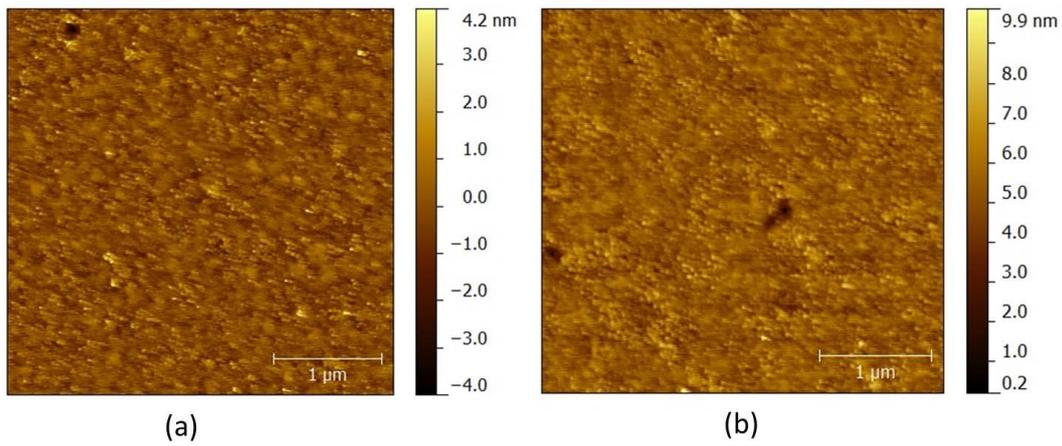

Fig. S3. Atomic force microscopy images of the sample surface (a) before gating experiment with RMS roughness of 0.57 nm and (b) after gating experiment with RMS roughness of 0.65 nm.

## 3. Estimation of induced carrier concentration

To characterize the field effect-induced carrier concentration, $n$, we use the term $n$

defined as electrons per formula unit and identify $n$ with $x$, the Ce doping level in $Pr_{2-x}Ce_xCuO_4$. Therefore, the optimally doped sample $Pr_{1.85}Ce_{0.15}CuO_4$ has a carrier concentration $x$=0.15 (equal to the Ce doping level) electrons per formula unit. Below we show an example on the estimation of $x$ at various $V_G$ of sample A. In order to estimate $x$ at every $V_G$, we make assumptions that $1/R_s(180K)$ is proportional to the $x$ [11, 12] and the charging state at $V_G$=3.0 V is optimally doped. As is shown in Fig. S4, $T_c$ increases and normal-state resistance decreases as the $V_G$ is increased from 2.67 to 3.0 V. However, at higher $V_G$ of 3.2 V, normal-state resistance keeps decreasing but $T_c$ decreases. Moreover, in the normal state, the $T$ dependence of $R_s$ shows the evolution from a metallic to insulating behavior with a minimum resistance $T$ ($T_{min}$) and $T_{min}$ keeps decreasing as the $V_G$ increase from 2.67 to 3.2 V, which is similar to bulk materials as the doping is increased. These indicate that the sample is overdoped at $V_G$=3.2 V. Therefore, it could be assumed that the sample at $V_G$=3.0 V is optimally doped, and thus, the carrier concentration is $x$=0.15 electrons per formula unit. The $x$ at other $V_G$ can be estimated by $x=0.15 \times R_s(x=0.15, 180K)/R_s(x, 180K)$. Here, we choose the $R_s$ at 180 K since at this $T$ the ionic liquid is frozen and a new charging state has already been established [11, 12]. The dependence of the estimated $x$ on the measured $R_s(180 K)$ is shown in Fig. S5. One can see that the modulation of $x$ from 0.0135 to 0.153 is obtained by IL-assisted field-effect doping.

Based on the estimated $x$, we can construct the phase diagram of $T_c$ versus $x$ (Fig. 3 in the main text). One can see that the superconductivity occurs at critical doping level of $x_c \approx 0.12$. Note that there has been a fair amount of controversy over the critical doping level at which superconductivity occurs [19, 27, 45-47]. The phase diagrams obtained by diverse groups using different methods show varied behavior. In our results the estimation of $x$ on three different EDLT devices is $x_c \approx 0.12$, consistent with the well-accepted previous results [19, 27, 46]. This can be seen from the consistency of phase diagrams induced by field effect doping and chemical doping (Fig. 3 in the main text). The origin of this controversy is still not understood. Nevertheless, the estimation of doping level does not affect the main conclusions of the current results.

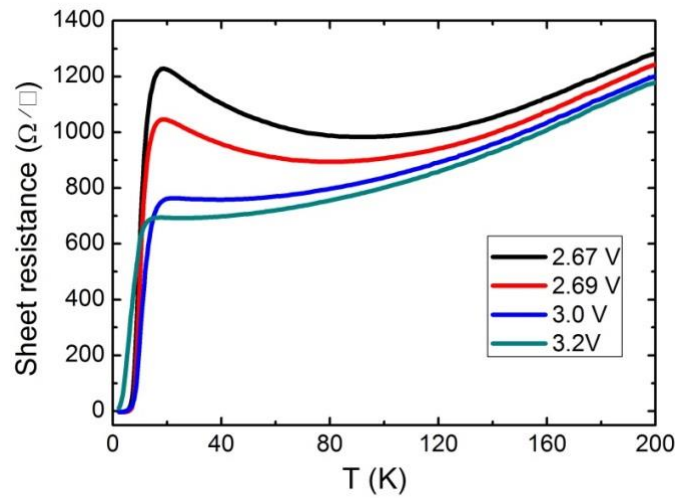

Fig. S4. Sheet resistance as a function of temperature at $V_G$ from 2.67 to 3.2 V.

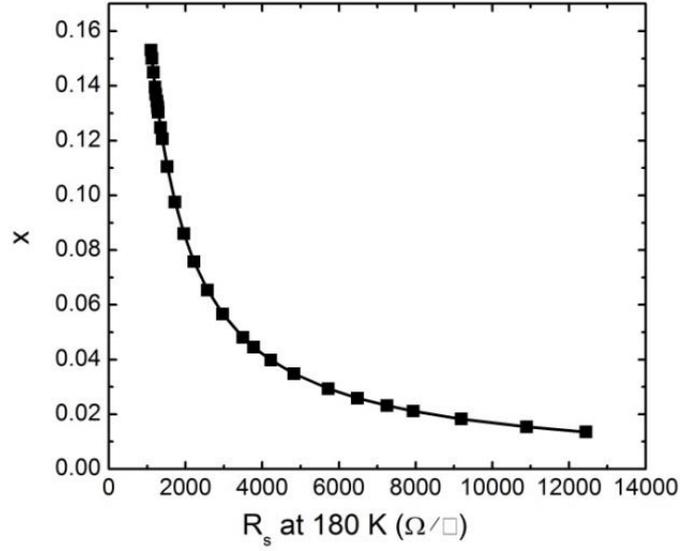

Fig. S5. The induced carrier concentration ($x$) as a function of the measured sheet resistance at 180 K.

## 4. Resistance under magnetic field

Magnetic field ($B$)-tuned SITs for sample A were done at three different $V_G$ of 2.75, 3.0 and 3.2 V, which correspond to under-, optimally- and over-doped states ($x$=0.132, 0.15 and 0.153), respectively. Figure S6(a) shows $R_s$-$T$ curves without $B$. The normal-state sheet resistance ($R_n$), obtained at the onset of superconductivity and zero $B$, are 1957, 764 and 694 Ω/□ for $V_G$=2.75, 3.0 and 3.2 V, respectively. Figure S6(b)-(d) are the $R_s$-$T$ curves at various $B$ for different $V_G$. The applied $B$ is perpendicular to the CuO$_2$ plane. It can be seen that at each $V_G$ the sample is initially superconducting and could be tuned to insulating states.

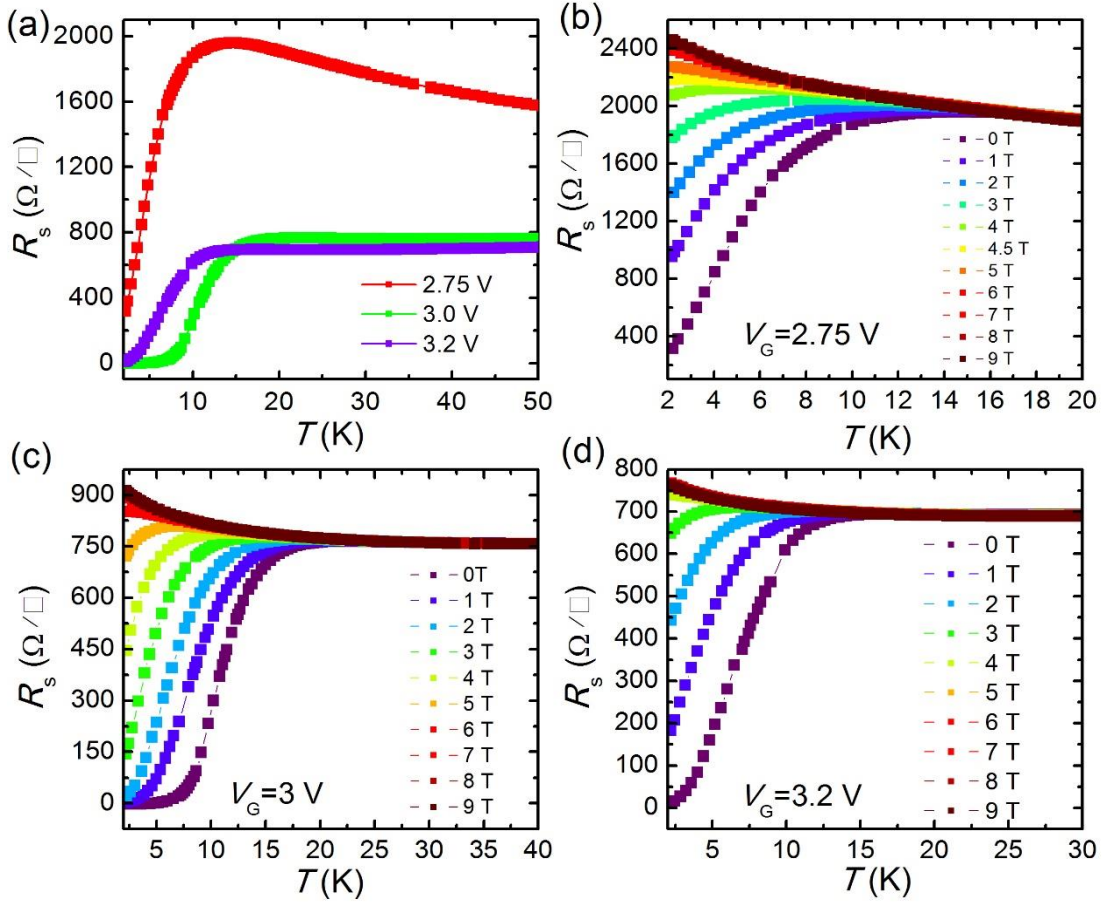

Fig. S6. (a) $R_s$-$T$ curves for sample A at $V_G$=2.75, 3.0 and 3.2 V, corresponding to the underdoped, optimally doped and overdoped states ($x$=0.132, 0.15 and 0.153), respectively. (b), (c), (d) are the $R_s$-$T$ curves at various magnetic field for different $V_G$.

## 5. Reproducibility of PCCO-EDLT experiments

Figure S7 shows $R_s$ isotherm curves at various $T$ for sample A and the inset shows the finite-size scaling analysis. Figure S8 shows $R_s$-$T$ curves at various $V_G$, $R_s$ isotherm curves and the finite-size scaling analysis for sample B. Figure S9 shows $R_s$-$T$ curves, $R_s$ isotherm curves and the scaling analysis of carrier-tuned SIT, $R_s$ isotherm curves and the scaling analysis of $B$-tuned SIT on another sample which is different from those in the main text. One can see that on those samples, the initially insulating states can be tuned into superconducting states by electric field effect. For carrier-tuned SITs, the obtained critical exponents $vz$ are similar to those of the samples in the main text and consistent to that in the quantum percolation model ($vz=7/3$). For the $B$-induced SIT (Fig. S9(c)), the finite-size scaling analysis gives $vz=1.2$. This value is slightly smaller than that of sample in the main text, but it is still close to that of the classical percolation model ($vz=4/3$). Moreover, the critical sheet resistances $R_c$ of these samples are comparable to those in the main text, and lower than $R_Q$. These results indicate that our results are reproducible.

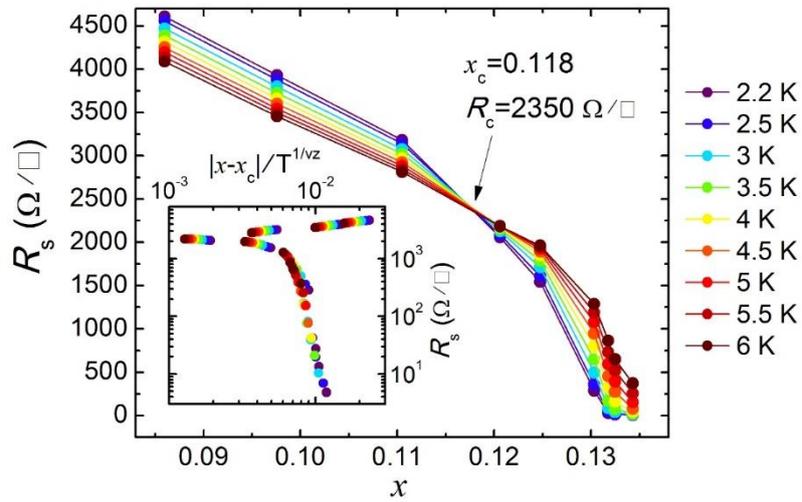

Fig. S7. Isotherms of $R_s$ as a function of $x$ at $T$ ranging from 2.2 to 6 K for sample A. The inset is the finite-size scaling of the isotherm curves, showing the resistance as a function of $|x-x_c|/T^{1/vz}$ with a critical exponent $vz=2.5$.

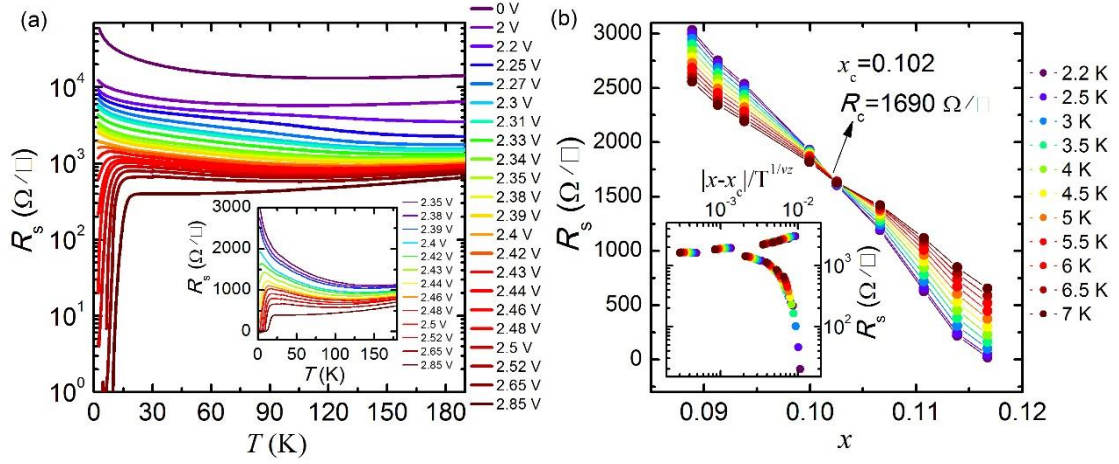

Fig. S8. (a) The logarithmic-scale $R_s$-$T$ curves at various $V_G$ for sample B. The inset of (a) is the linear-scale $R_s$-$T$ curves near the SIT. (b) Isotherms of $R_s$ as a function of $x$ at $T$ ranging from 2.2 to 7 K. The inset of (b) is the finite-size scaling of the isotherm curves, showing the resistance as a function of $|x-x_c|/T^{1/vz}$ with a critical exponent $vz$=2.4.

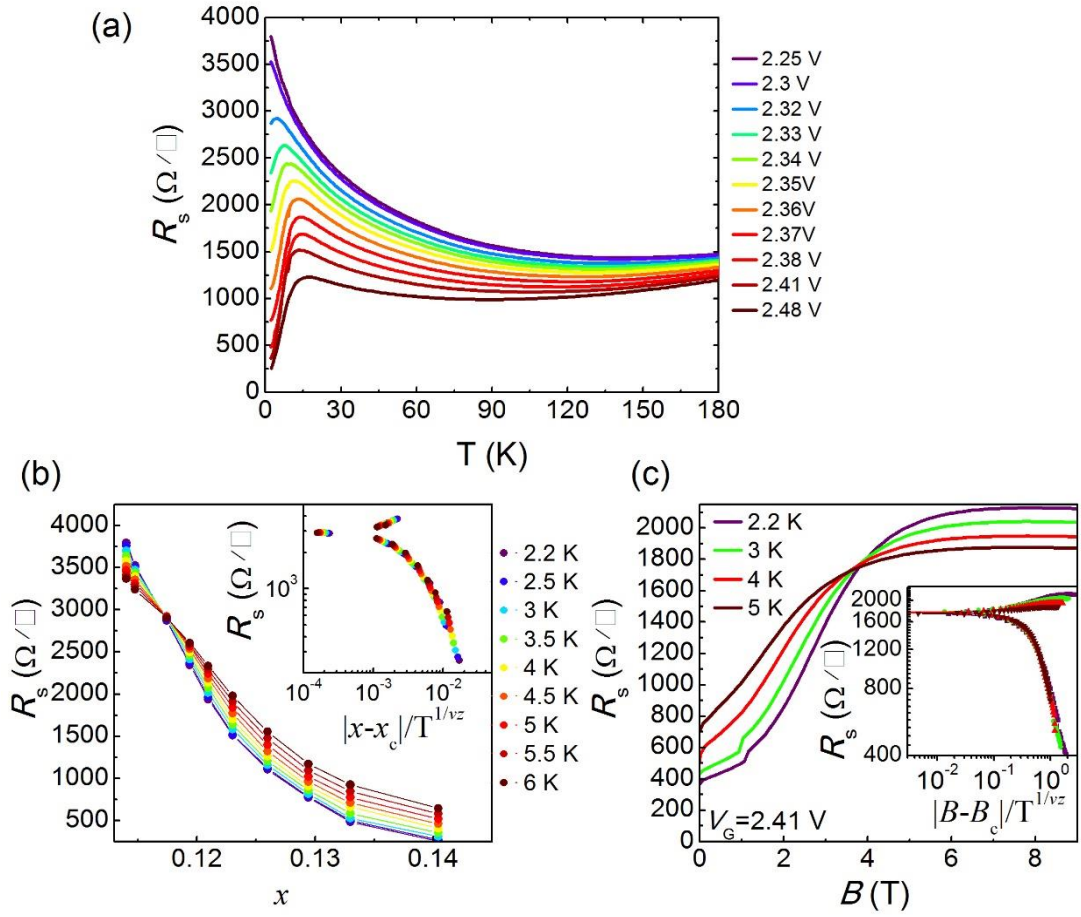

Fig. S9. (a) $R_s$-$T$ curves near the SIT for another sample. (b) $R_s$ isotherms as a function of $x$ at $T$ ranging from 2.2 to 6 K. The inset is the finite-size scaling of the isotherm curves, showing the resistance as a function of $|x-x_c|/T^{1/vz}$ with a critical exponent $vz$=2.5. (c) $R_s$ as a function of magnetic field at four different $T$ at $V_G$=2.41 V. The inset shows the $R_s$ as a function of $|B-B_c|/T^{1/vz}$ with $vz$=1.2.

## 6. Insulating behavior near critical point

At the doping level near the critical point $x_c$ where superconductivity occurs, with decreasing $T$ the resistances of cuprates show metallic behavior and then insulating behavior below certain $T$ ($T_{min}$) at which normal-state resistance shows a minimum value. In chemically doped cuprates, the resistance upturn at low $T$ could be attributed to various effects, for example, weak localization [48], Kondo scattering [49, 50] or spin scattering [51, 52]. Moreover, the $T_{min}$ of the chemically doped insulating samples is normally below ~70 K [48, 51, 53, 54]. However, the normal-state resistance behavior of PCCO-EDLT in the present results is different from that in chemically doped samples. At the insulating phase near $x_c$ the $T_{min}$ is higher than ~140 K in PCCO-EDLT, as is shown by the black dots in Fig. S10 for samples A (other samples show similar $T_{min}$ near $x_c$). These suggest that at the insulating phase near $x_c$, PCCO-EDLT show much higher insulating state compared with the chemically doped samples. Moreover, for electron-doped cuprates, it is suggested that the antiferromagnetic state can persist up to the optimally doped level [19, 55], the insulating behavior at low $T$ can also be due to the antiferromagnetic order. Therefore, since in current PCCO-EDLT, the resistance shows upturn at high $T$, it is believed that these films show the resistance upturn primarily because they are insulators.

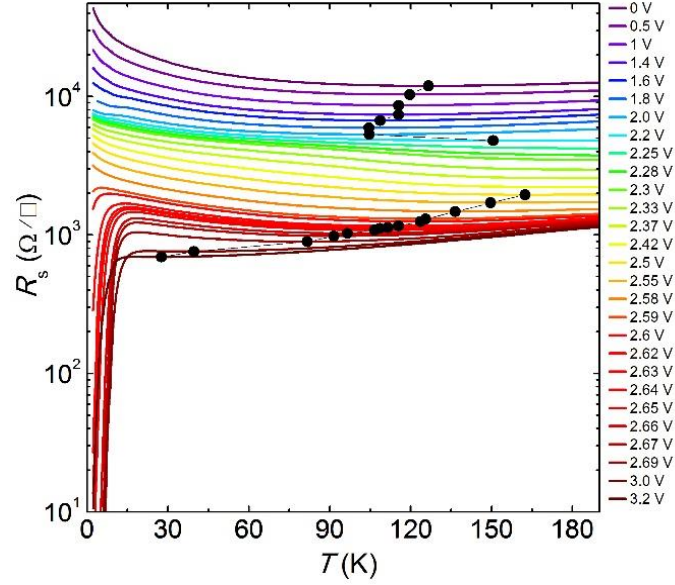

Fig. S10. $R_s$-$T$ curves at different $V_G$ of sample A. The black solid dots are the $T_{min}$'s at which the sample shows a minimum resistance at each curve. At temperatures below $T_{min}$, the sample show insulating behavior.

## 7. Hall-effect measurement

It has been observed that QPTs occur inside the superconducting dome, which were evidenced by the observation of a Hall coefficient anomaly appear near optimum doping level [56, 57]. We also measured Hall effect and show normalized carrier density $n_H$ as a function of doping level in Fig. S11. One can see that there is an anomaly in $n_H$ as $x$ is increased across $x \sim 0.13$. $n_H$ increases moderately at $0.02<x<0.13$, and sharply at $0.13<x<0.15$. However, since the overdoped state could not be obtained in the present experiment, we could not observe the peak of $n_H$ near optimally doped state which suggests QPT [56, 57]. Whether the QPT can be probed by Hall effect on the PCCO-EDLT is not clear now.

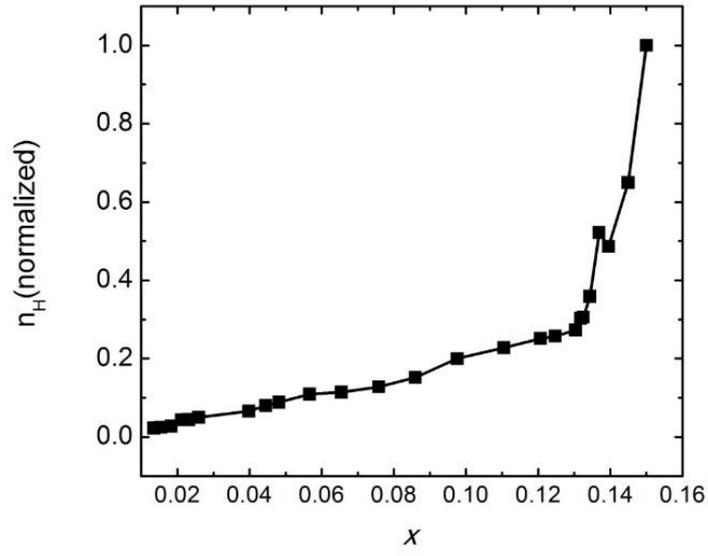

Fig. S11. Normalized carrier density $n_H$ (measured from Hall effect) as a function of doping level.